\documentclass[twoside,twocolumn,9pt]{article}
\usepackage{extsizes}
\usepackage[super,sort&compress,comma]{natbib} 
\usepackage[version=3]{mhchem}
\usepackage[left=1.5cm, right=1.5cm, top=1.785cm, bottom=2.0cm]{geometry}
\usepackage{balance}
\usepackage{times,mathptmx}
\usepackage{sectsty}
\usepackage{graphicx} 
\usepackage{lastpage}
\usepackage[format=plain,justification=justified,singlelinecheck=false,font={stretch=1.125,small,sf},labelfont=bf,labelsep=space]{caption}
\usepackage{float}
\usepackage{fancyhdr}
\usepackage{fnpos}
\usepackage[english]{babel}
\usepackage{array}
\usepackage{droidsans}
\usepackage{charter}
\usepackage[T1]{fontenc}
\usepackage[usenames,dvipsnames]{xcolor}
\usepackage{setspace}
\usepackage[compact]{titlesec}

\usepackage{blindtext}

\definecolor{cream}{RGB}{222,217,201}

\definecolor{N2}{HTML}{373640}
\definecolor{N3}{HTML}{63686e}
\definecolor{N4}{HTML}{7e97a6}
\definecolor{N5}{HTML}{adccc7}
\definecolor{N6}{HTML}{0c907d}
\definecolor{N7}{HTML}{b6f7c1}
\definecolor{N8}{HTML}{144c52}
\definecolor{N9}{HTML}{053220}
\definecolor{N10}{HTML}{133353}
\definecolor{N11}{HTML}{4A6B8A}
\definecolor{N12}{HTML}{000000}

\definecolor{NT}{HTML}{808080}
\definecolor{IN}{HTML}{484848}

\newcommand\crule[3][black]{\textcolor{#1}{\rule{0.7cm}{0.25cm}}}

\begin{document}

\pagestyle{fancy}
\thispagestyle{plain}
\fancypagestyle{plain}{

\renewcommand{\headrulewidth}{0pt}
}

\makeFNbottom
\makeatletter
\renewcommand\LARGE{\@setfontsize\LARGE{15pt}{17}}
\renewcommand\Large{\@setfontsize\Large{12pt}{14}}
\renewcommand\large{\@setfontsize\large{10pt}{12}}
\renewcommand\footnotesize{\@setfontsize\footnotesize{7pt}{10}}
\makeatother

\renewcommand{\thefootnote}{\fnsymbol{footnote}}
\renewcommand\footnoterule{\vspace*{1pt}%
\color{cream}\hrule width 3.5in height 0.4pt \color{black}\vspace*{5pt}} 
\setcounter{secnumdepth}{5}

\makeatletter 
\renewcommand\@biblabel[1]{#1}            
\renewcommand\@makefntext[1]%
{\noindent\makebox[0pt][r]{\@thefnmark\,}#1}
\makeatother 
\renewcommand{\figurename}{\small{Fig.}~}
\sectionfont{\sffamily\Large}
\subsectionfont{\normalsize}
\subsubsectionfont{\bf}
\setstretch{1.125} 
\setlength{\skip\footins}{0.8cm}
\setlength{\footnotesep}{0.25cm}
\setlength{\jot}{10pt}
\titlespacing*{\section}{0pt}{4pt}{4pt}
\titlespacing*{\subsection}{0pt}{15pt}{1pt}

\fancyfoot{}
\fancyfoot[RO]{\footnotesize{\sffamily{1--\pageref{LastPage} ~\textbar  \hspace{2pt}\thepage}}}
\fancyfoot[LE]{\footnotesize{\sffamily{\thepage~\textbar\hspace{3.45cm} 1--\pageref{LastPage}}}}
\fancyhead{}
\renewcommand{\headrulewidth}{0pt} 
\renewcommand{\footrulewidth}{0pt}
\setlength{\arrayrulewidth}{1pt}
\setlength{\columnsep}{6.5mm}
\setlength\bibsep{1pt}

\makeatletter 
\newlength{\figrulesep} 
\setlength{\figrulesep}{0.5\textfloatsep} 

\newcommand{\topfigrule}{\vspace*{-1pt}%
\noindent{\color{cream}\rule[-\figrulesep]{\columnwidth}{1.5pt}} }

\newcommand{\botfigrule}{\vspace*{-2pt}%
\noindent{\color{cream}\rule[\figrulesep]{\columnwidth}{1.5pt}} }

\newcommand{\dblfigrule}{\vspace*{-1pt}%
\noindent{\color{cream}\rule[-\figrulesep]{\textwidth}{1.5pt}} }

\makeatother

\twocolumn[
  \begin{@twocolumnfalse}
\vspace{3cm}
\sffamily
\begin{tabular}{m{4.5cm} p{13.5cm} }

 & \noindent\LARGE{\textbf{Patchy particles by self-assembly of star copolymers on a spherical substrate: Thomson solutions in a geometric problem with a color constraint}} \\
\vspace{0.3cm} & \vspace{0.3cm} \\

 & \noindent\large{Tobias M. Hain,\textit{$^{a,c}$}, Gerd E. Schr\"oder-Turk$^{\ast}$\textit{$^{a,b,c}$} and Jacob J. K. Kirkensgaard$^{\ast}$\textit{$^{b}$}} \\

 & \noindent\normalsize{ Confinement
  or geometric frustration is known to alter the structure of soft
  matter, including copolymeric melts, and can consequently be used to
  tune structure and properties. Here we investigate the self-assembly
  of $ABC$ and $ABB$ 3-miktoarm star copolymers confined to a
  spherical shell using coarse-grained Dissipative Particle Dynamics
  simulations. In bulk and flat geometries the $ABC$ stars form
  hexagonal tilings, but this is topologically prohibited in a
  spherical geometry which normally is alleviated by forming
  pentagonal tiles. However, the molecular architecture of the $ABC$
  stars implies an additional 'color constraint' which only allows
  even tilings (where all polygons have an even number of edges) and
  we study the effect of these simultaneous constraints.  We find that
  both $ABC$ and $ABB$ systems form spherical tiling patterns, the
  type of which depends on the radius of the spherical substrate. For
  small spherical substrates, all solutions correspond to patterns
  solving the Thomson problem of placing mobile repulsive electric
  charges on a sphere.  In $ABC$ systems we find three coexisting,
  possibly different tilings, one in each color, each of them solving
  the Thomson problem simultaneously. For all except the smallest
  substrates, we find competing solutions with seemingly degenerate
  free energies that occur with different
  probabilities. Statistically, an observer who is blind to the
  differences between $B$ and $C$ can tell from the structure of the
  $A$ domains if the system is an $ABC$ or an $ABB$ star copolymer
  system. }\\

\end{tabular}

 \end{@twocolumnfalse} \vspace{0.6cm}

  ]

\renewcommand*\rmdefault{bch}\normalfont\upshape
\rmfamily
\section*{}
\vspace{-1cm}


\footnotetext{\textit{$^{a}$~  College of Science, Health, Engineering and Education, Mathematics and Statistics, Murdoch University, 
90 South Street, 6150 Murdoch, Western Australia, Australia}}
\footnotetext{\textit{$^{b}$~Department of Food Science, University of Copenhagen, Rolighedsvej 26, 1958 Frederiksberg, Copenhagen, Denmark }}
\footnotetext{\textit{$^{c}$~Physical Chemistry, Department of Chemistry, Lund University, P.O. Box 124, 221 00 Lund, Sweden}}
\footnotetext{$^{\ast}$~G.Schroeder-Turk@murdoch.edu.au,~jjkk@food.ku.dk}





\section{Introduction}

\begin{figure}
  \center \includegraphics[width=.4\textwidth]{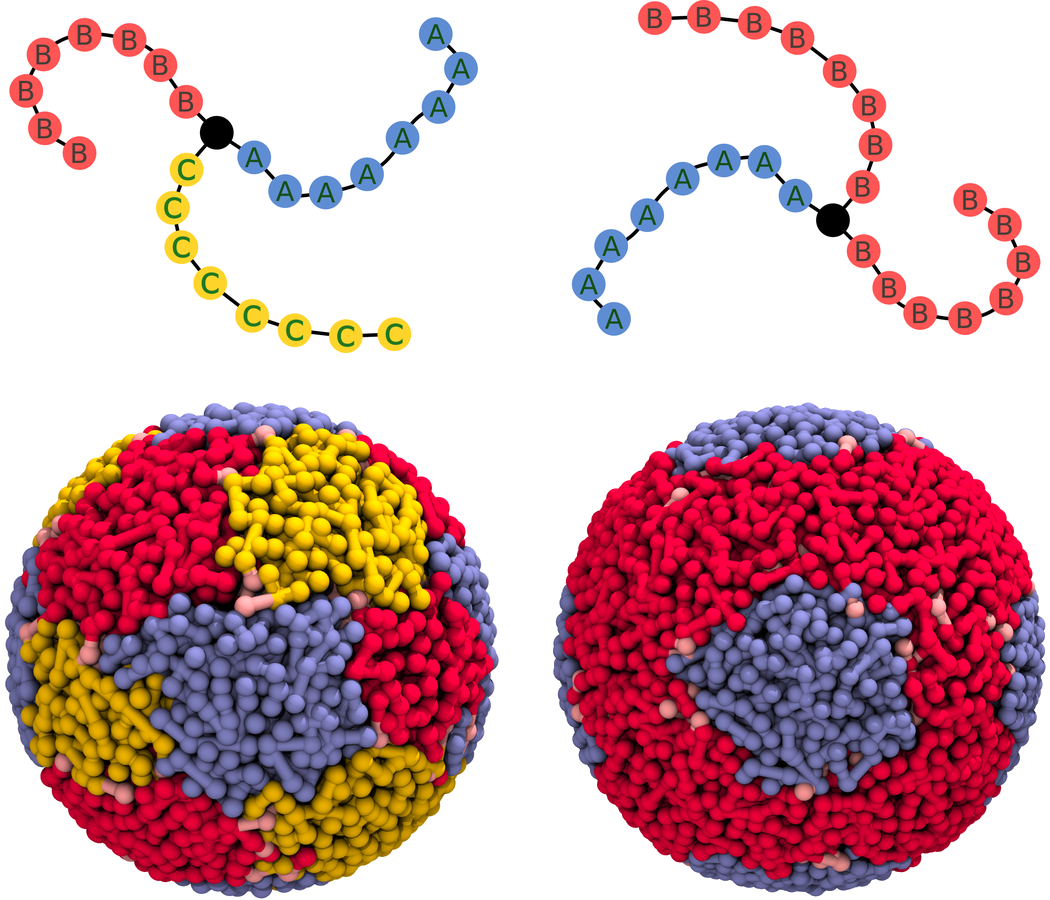}
  \caption{\textbf{Polymeric self-assembly of $ABC$ and $ABB$ star
      copolymers on a spherical substrate} \textit{Top panel:}
    Schematic visualization of coarse grained models of $ABC$ and
    $ABB$ star copolymers used for the DPD simulations. \textit{Bottom
      panel:} Snapshots of simulations comprising $ABC$ and $ABB$ star
    copolymers. Polymer arms of identical color agglomerate into
    patches. The $ABC$ system creates a three colored tiling of the
    sphere, whereas the $ABB$ system builds a single tiling made up of
    only $A$-type patches in a $B$-type matrix.}
  \label{fig:ABC-polymers}
\end{figure}

\begin{figure}
  \center \includegraphics[width=.45\textwidth]{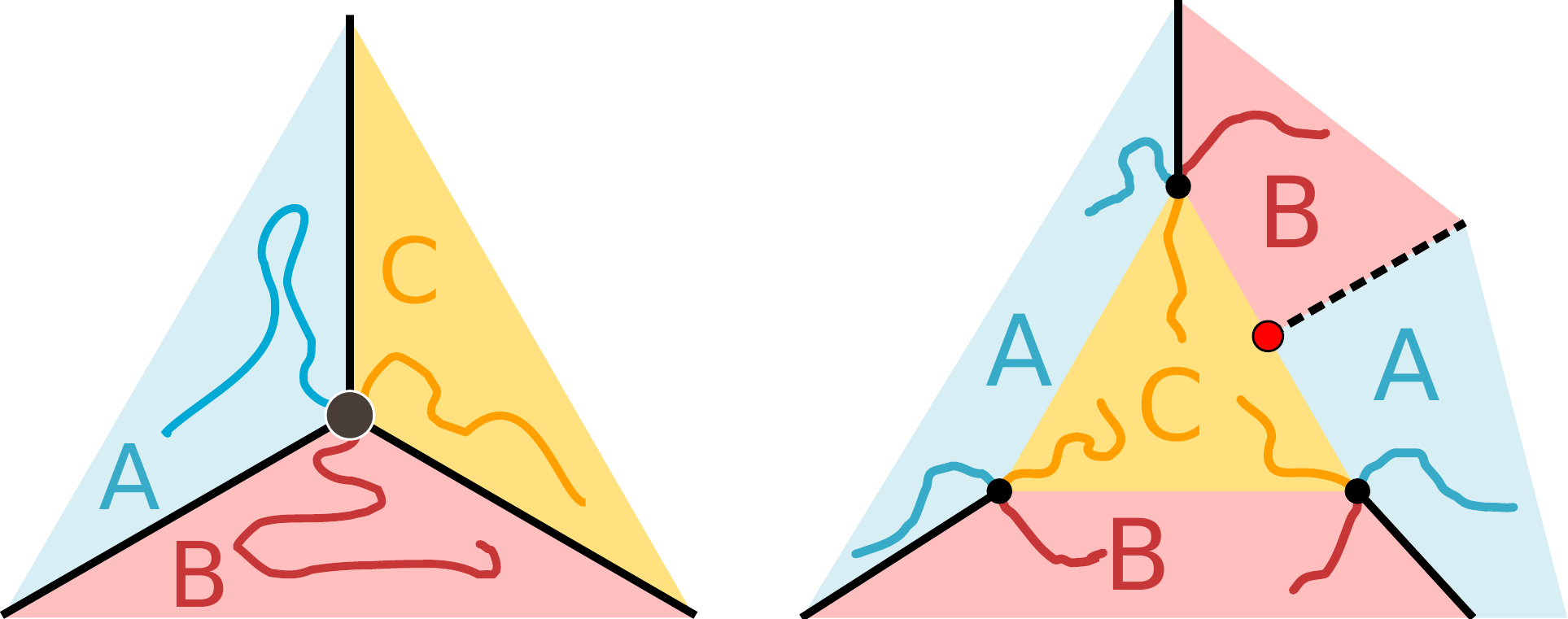}
  \caption{ \textbf{Structural constraints imposed by polymer
      architecture: the color-constraint} \textit{Left panel:} Since
    the three different polymer arms making up domains are bonded at a
    central junction bead (grafting point), the latter must sit on points or lines
    where three different domains meet so that each arm may extend in
    a domain of its species. \textit{Right panel:} For tiling patterns, this
    results in the so-called \textit{color constraint}: only polygons
    with an even number of edges are allowed, where the types of all
    adjacent polygons alternate. The figure illustrates this: if a
    polygon with an uneven number of edges is attempted to be formed,
    a new interface (dashed line) and grafting point (red point) is
    introduced by the architecture of the stars, resulting in an even
    polygon.}
  \label{fig:color-constraint}
\end{figure}

The self-assembly of linear diblock copolymers and their phase diagram
is nowadays well understood \citep{bates2005, matsen2007,
  grason2004}. By contrast, the study of the phase behaviour of more
complex copolymer architectures, like grafts or stars
\citep{polymeropoulos2017}, remain incomplete, due to the larger
parameter space and, hence, a larger variety of possible structures
\citep{matsen2007, grason2004, guo2008, tyler2007, fischer2014}.  Here
we consider $ABC$ 3-miktoarm star terpolymers, henceforth called $ABC$
star copolymers. These are copolymers which consist of three linear
chains connected at a central grafting point \citep{okomato1997,
  kirkensgaard2009, grason2004, polymeropoulos2017}, as shown in
fig.~(\ref{fig:ABC-polymers}) or (\ref{fig:color-constraint}). These
star copolymers can be synthesized so that the three arms are
immiscible; herein we refer to the three polymeric species as colors:
blue, yellow and red. When this immiscibility drives the moieties to
micro-phase separation, the arms of equal species will agglomerate
into domains, which will self-assemble into complex
structures. Considerable amount of work has been put towards the
investigation of one type of these structures: columnar phases whose
cross-sections are planar tiling patterns \citep{kirkensgaard2014a,
  kirkensgaard2012b, gemma2002, decampo2011, hayashida2007,
  hayashida2006, takano2004, takano2005, tang2004, zhang2010,
  matsushita2007}.
A distinguished and important feature of $ABC$ star copolymers is
that all structures arising from these molecules must be compatible
with the special architecture of the latter: 

\begin{enumerate}
\item The grafting points form triple lines\citep{decampo2011} where
  three different domains meet, which, in cross-section, corresponds
  to vertices of the tiling pattern.
\item Any given patch of a given color (e.g. yellow), must be
  surrounded by an alternating sequence of patches of the other colors
  (e.g. blue and red). The number of these surrounding patches must
  then be even \citep{gemma2002}.
\end{enumerate}

\noindent For more details see fig.~(\ref{fig:color-constraint}). In
this article we will refer to constraint 2 as the
\textit{color-constraint}.

If the star copolymers are chosen symmetrically, i.e.~all arms have
equal length, and with equal interaction strength between the arms, a
hexagonal columnar phase is formed where a cross section perpendicular
to the columns yields a planar 3-colored honeycomb pattern which we
here consider as the 'ground state' of the system.
\citep{kirkensgaard2012a, kirkensgaard2012b, kirkensgaard2014a,
  gemma2002, decampo2011, zhang2010, takano2004}. The hexagonal tiling
can be tuned into a large variety of tilings by varying the length of
one of the three arms \citep{kirkensgaard2014a, gemma2002,
  zhang2010}. However, all these tilings consist of vertices of order
three only, which is enforced by the molecular architecture of the
star copolymers. 

Apart from changing the chemical composition or interactions of the
polymers, another way to tune structures is by geometric
confinement. A simple analogy illustrates this fundamental geometric
concept: the peel of an orange for example cannot be confined to a
flat plane, without tearing or deforming it. This also applies for the
star polymers: the optimal free energy configuration they form in the
plane, the regular honeycomb, cannot be fitted on a spherical
substrate without distorting the planar pattern.

Unlike the restriction of polymers to a thin film, the confinement to
curved geometries, like spheres, does not only impose the constraint
of physical confinement onto the polymers, but also introduces
curvature to the system. This alters the shape and structure of the
space available to the polymer melt which can enforce or prohibit some
structures to form. Such curvature-related effects have been described
for multiple self-assembly systems:

Several articles report on the influence of curvature on hexagonal
particle orders on surfaces with positive and negative curvature, both
using experiments \cite{irvine2010, irvine2012, guerra2018,
  lipowsky2005a, lipowsky2005b, bausch2003} and simulations
\citep{giarritta1992, giarritta1993, guerra2018}. Two dimensional tilings can be
created from these particle assemblies by assigning each particle a
polygon where the number of edges coincides with its coordination
number, which is the number of neighbouring particles. While these
particles would arrange in a hexagonal order in a plane, and
therefore, form a perfect hexagonal tiling, defects in this patterns
were found after the particles self-assembled on curved surfaces.
Zhang \textit{et~al.} \cite{zhang2014} found similar defects in the
self-assembly of $AB$ diblock copolymers confined to a spherical
substrate using numerical methods to solve the Landau-Brazovskii
theory. For cylinder forming diblocks, the cylinders distributed over
the surface of the sphere in a generally hexagonal order, however,
5-fold defects were found. For larger systems, scars of connected 5-
and 7-fold defects occur.  These defects have a fundamental
mathematical origin: the different topologies of the confining
surface. Each tiling and polyhedra (and topological equivalents) have
an intrinsic property, the Euler characteristic $\chi$, describing its
topological type \citep{kamien2002, bowick2009, conway2016}. An Euler
characteristic of $\chi=2$ corresponds to an object that is a single
component without any handles or cavities, such as the sphere.  A
given tiling can only tessellate a surface of the same topological
space, therefore surfaces having the same Euler characteristic as the
tiling\cite{bowick2009}. A planar, periodical hexagonal tiling has
$\chi=0$, as does a torus. Therefore the hexagonal tiling can be
mapped onto the latter. When a hexagonal lattice is forced onto an
incompatible curved surface, as for example a sphere with $\chi=2$,
the mismatch leads to 'geometric frustration': the hexagonal lattice
is incompatible with the topology of the substrate. To cope with this
incompatibility defects occur in the hexagonal order.

To check if a tiling is compatible with a sphere, Euler's formula can
be used, which reads in case of a sphere\citep{kamien2002,
  conway2016}:
\begin{equation}
  \chi = V - E + F = 2
  \label{eq:eulers-formula}
\end{equation}

where $V$, $E$, $F$ is the number of vertices, edges and facets in the
tiling. If a tiling fulfills this condition, it can be mapped onto a
sphere without defects. In our case, where tilings are generated by
$ABC$ star copolymers, the color constraint can be incorporated into
eq.~(\ref{eq:eulers-formula}). Since only vertices where three edges
meet are allowed, each edge is shared by two and each vertex by three
facets (see also fig.~\ref{fig:color-constraint}). In this case,
eq.~\ref{eq:eulers-formula} can then be expressed in terms of the
number of polygons in the tiling:

\begin{equation}
  \chi = \sum_i \left( \frac{i \cdot n_i}{3} - \frac{i \cdot n_i}{2} + n_i \right) = \sum_i \left( 1 - \frac{1}{6}i \right) n_i = 2
  \label{eq:euler-colorconstraint}
\end{equation}
where $n_i$ is the number of polygons with $i$ edges in the
tiling. This equation easily shows that a hexagonal tiling is
incompatible, since the term in the brackets equals zero for
$i=6$. Therefore polygons with a different number of edges need to be
introduced to fulfil the equation. One solution to
eq.~(\ref{eq:eulers-formula}) for the sphere is the arrangement of 12
pentagons to an icosahedron. Since the left side of
eq.~(\ref{eq:eulers-formula}) vanishes for hexagons, an arbitrary
number of the latter can be added to the 12 pentagons and the topology
will not change. A well-known configuration is the soccer ball,
consisting of 12 pentagons and 20 hexagons.
Apart from investigations of particles from polymer systems, abstract
systems with topological defects were investigated analytically
\citep{bowick2000, bowick2002, bowick2009, bowick2011}. Here the
behavior of abstract disclinations from the crystalline state,
for example particles with 5 neighbours in an otherwise hexagonal
lattice, was investigated using free energy calculations. The results
agree with the results found in the physical particle and polymer
systems: the favoured state are 12 5-fold disclinations, also
the above mentioned scars (connected disclinations) are found.
 
A very prominent problem of ordering on a sphere is the so-called
Thomson problem. It was formulated by J.~J.~Thomson in 1904 in the
context of his atomic model. The Thomson problem is the search for the
minimal energy configuration of $n$ repelling electrons, all of the
same negative charge $-e$, on a spherical
surface\citep{thomson1904}. The resulting arrangement of electrons and
their symmetries
\citep{erber1991,bondarenko2015,altschuler1997,wales2006,wales2009}
has been found in many seemingly unconnected problems, as for example
in the design of protein virus capsides
\citep{caspar1962,mannige2008,rochal2016}, the construction of
fullerens and nanotubes \cite{robinson2013}, but also in more
generalized Thomson problem versions \citep{mughal2014}. To reach the
minimal energy solution, the optimal coordination number of a single
electron is six, however, due to the geometric frustration defects in
the hexagonal order must occur, as explained above
\citep{garrido1997}.  For our system, it is useful to interpret the
electron positions of the Thomson problem solutions as vertices of a
polyhedra. The graph of its dual polyhedron is a tiling of the sphere,
where each electron is assigned a tile whose number of edges is
equivalent to the coordination number of the corresponding electron. A
solution of the Thomson problem, henceforth called a Thomson
solutions, can therefore be described and labeled by its dual lattice,
see table~(\ref{tab:thomson-tilings}).

In conclusion, using $ABC$ star copolymers confined to a spherical
shell as a model system enables the simultaneous study of two
different constraints: geometric frustration and the influence of the
color-constraint. To investigate the effects of each on their own, a
strategy is needed to switch one of them on and off. This is
accomplished by using two different kind of star molecules, the
aforementioned $ABC$ stars and $ABB$ star molecules, see right panel
in fig.~(\ref{fig:ABC-polymers}). These only differ to the $ABC$ stars
in that two arms are of the same species. Thus the color-constraint
can be eliminated, since the grafting points of the $ABB$ stars can
move freely across the interface between $A$ and $B$ type domains. The
$A$ type domains, which are the tiles in the resulting tiling, can
then freely move around in a $B$ type matrix.

\section{Methods}
\label{sec:methods}

\subsection{Dissipative particle dynamics of star copolymers}

Dissipative Particle Dynamic (DPD) simulations are used to find
equilibrium configurations of the polymer systems. DPD simulations
\citep{hoogerbrugge1992, espanol1995} are a type of molecular dynamic
simulations designed for coarse grained models of molecules, which
makes it a natural fit for polymer melts \citep{kirkensgaard2010,
  kirkensgaard2011,kirkensgaard2012a,kirkensgaard2012b}. As all
molecular dynamics simulations, the DPD method is based on the forward
integration of Newton's equation of motion in time for each particle
$i$:
\begin{equation*}
  \frac{d^2 \textbf{x}_i}{dt^2} = \frac{1}{m} \cdot \textbf{F}_i
\end{equation*}
In our case, a particle is a single bead in the polymer arms
(see fig.~(\ref{fig:ABC-polymers})), where each bead may represent
many atoms. A symmetric star copolymer then consists of a center
particle with three connected arms, each consisting of a chain of
bonded particles. A schematic representation of such a coarse grained
polymer is shown in fig.~(\ref{fig:ABC-polymers}).

We use the simulation package \textsc{hoomd-blue} \cite{anderson2008,
  glaser2015, phillips2011} to perform our simulations. We will only
briefly discuss the parameters used at this point, for details on the
implementation we refer to \citep{groot1997} and the documentation of
the \textsc{hoomd-blue} package \citep{hoomd-url}. In this simulation
package all units are given based on three reference units (distance
$\mathcal{D}$, energy $\mathcal{E}$ and mass $\mathcal{M}$) which can
be chosen arbitrarily. All other units, for example a force, can be
derived from these units, for more details we refer to the
\textsc{hoomd-blue} manual\citep{hoomd-url}. In the course of this
article, all given values are given in terms of these reference units
unless stated otherwise. The package implements the DPD method
following the formulation of \cite{groot1997,espanol1995}. Here the
force on particle $i$ is given as
\begin{equation*}
\textbf{F}_i=\sum_{i \neq j} \left( \textbf{F}_{ij}^{\text{C}} +
\textbf{F}_{ij}^{\text{D}} + \textbf{F}_{ij}^{\text{R}} \right)
\label{eq:dpd-force}
\end{equation*}
where the sum is over all particle pairs within a cutoff radius
$r_c=1$ around the $i$-th particle. The force consists of three
contributions: a conservative force $\textbf{F}_{ij}^{\text{C}}$
representing the repulsive interactions between the particles, a
dissipative force $\textbf{F}_{ij}^{\text{D}}$ and a random force
$\textbf{F}_{ij}^{\text{R}}$. The latter two act as a thermostat to
keep the temperature of the system constant. Since a thermostat is a
built-in feature of the DPD interactions, the system is technically
advanced as a $NVE$ ensemble using a standard Velocity-Verlet step
algorithm, although it is effectively a $NVT$ ensemble. The
conservative force is 0 only for $r_{ij} \geq r_c$ and is otherwise
given by
\begin{equation}
  \textbf{F}_{ij}^{\text{C}}=
      a_{ij} \left( 1-\frac{r_{ij}}{r_c} \right) \hat{\textbf{r}}_{ij}
\end{equation}
where $a_{ij}$ is the maximum repulsion between two particles and
therefore a measure of the interactions strength, $\textbf{r}_{ij} =
\textbf{r}_{i} - \textbf{r}_{j}$ and $\hat{\textbf{r}}_{ij} =
\textbf{r}_{ij} / |\textbf{r}_{ij}|$. The interactions between two
particles of the same species is given as $a_{ii} = 75
\frac{k_{\text{B}} T}{\rho}$, where $\rho$ is the number density and
$k_{\text{B}} T$ the temperature in the polymer melt. The interaction
parameters can be mapped onto the well established Flory-Huggins
interaction parameter $\chi_{ij}$ \citep{flory1942, huggins1942} used
in polymer science using $a_{ij}=a_{ii} + 3.268 \chi_{ij}$
\citep{groot1998}. We use values of $a_{ii}=25$ and $a_{ij}=40$ that,
at a temperature of $k_B T = 1$ and a particle density of $\rho=3$,
corresponds to the strong segregation limit. The single beads of the
polymer chains are bonded by a harmonic potential, given as
$V_{\text{H}}(r) = \frac{1}{2} k (r-r_0)^2$, where $k$ measures the
strength of the bond and $r_0$ the bond rest length. In our system we
chose $k=4\,\frac{\mathcal{E}}{\mathcal{D}^2}$ and
$r_0=0.88\,\mathcal{D}$ as the position of the first peak of the pair
correlation function in a system of unbonded, identical particles with
the given interaction parameters. Each arm in the polymers consists of
8 beads.

\begin{figure}
  \hfill \includegraphics[width=.22\textwidth]{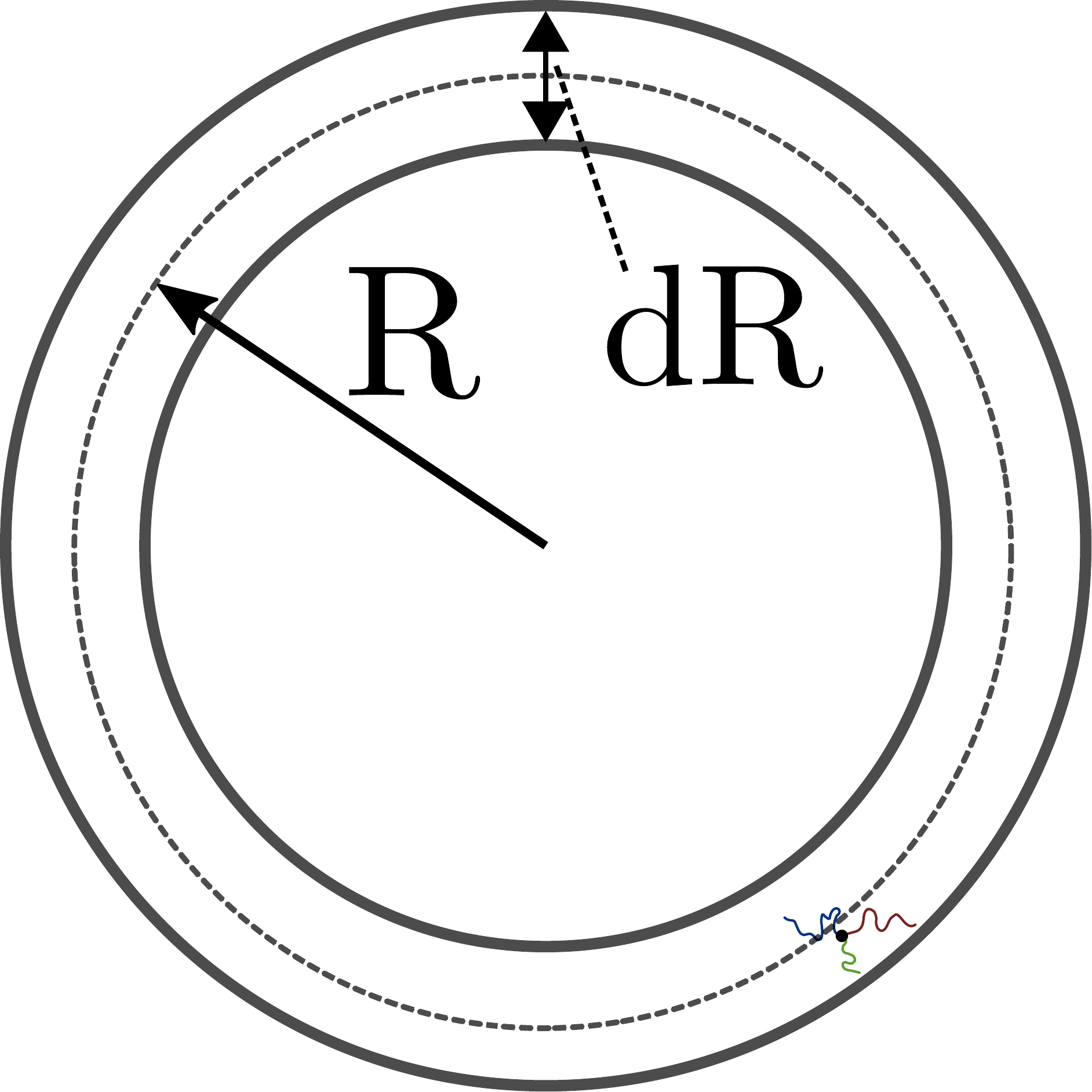} \hfill
  \raisebox{0.3\height}{\includegraphics[width=.25\textwidth]{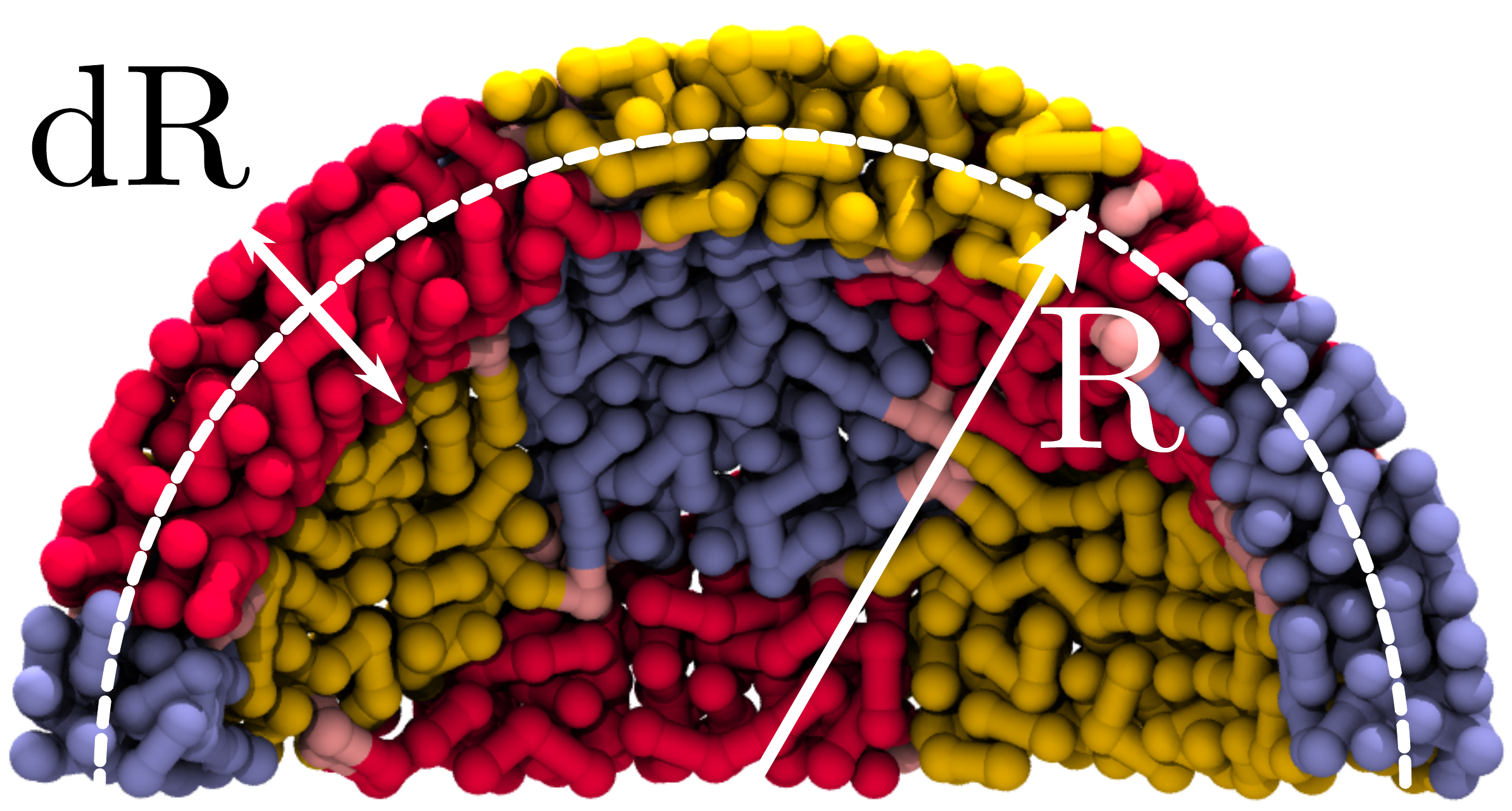} \hfill }
  \caption{\textbf{Simulation setup and substrate model} \textit{Left
      panel:} Schematic sketch of a cross section of the simulation
    setup. The two black, bold circles represent spherical, repulsive
    `Lennard-Jones walls' with radii of $R+\frac{dR}{2}+\sigma$ and
    $R-\frac{dR}{2}-\sigma$. As shown, the outer wall exerts a
    LJ-force on a particle along a vector perpendicular to the wall
    towards its center. The force of the inner shell acts
    outwards. Together these walls confine the polymers to a shell of
    thickness $dR$. \textit{Right panel:} Cross section through a
    simulation snapshot with radius $R=8\,\mathcal{D}$ with the same
    quantities marked. The image shows the homogeneity of structure in
    the radial direction of the shell and gives an idea for the scale
    of the system.}
  \label{fig:simbox-sketch}
\end{figure}

The confinement of the system to a spherical substrate is modelled as
follows: the simulation volume is a spherical shell bounded by two
repulsive spherical walls interacting with the polymers with a
purely repulsive Lennard-Jones potential:
\begin{equation*}
  V_{\text{LJ}}(r) = 4 \epsilon \left[ \left( \frac{\sigma}{r} \right)^{12} - \left( \frac{\sigma}{r} \right)^6 \right] + \Delta V
\end{equation*}
where $r$ in this case is the length of the vector from the particle
perpendicular to the wall, not to be confused with $r_{ij}$, the
pairwise distance between two particles, see
fig.~(\ref{fig:simbox-sketch}). $\Delta V = - \left( r-r_{\text{cut}}
\right) \frac{\partial V}{\partial r} \left( r_c \right) -
V_{\text{LJ}} \left( r_c \right)$ and $\sigma$ is the range of the
repulsive potential, $\epsilon$ would be the strength of the
attractive part of the Lennard-Jones potential, however, is of no
relevance in the purely repulsive version used here. While the outer
wall exerts a force towards its center, the forces of the inner wall
acts outwards. Hence, the wall keeps all particle inside the
spherical shell they enclose. We choose $\sigma=1\mathcal{D}$ and set
the cutoff of the wall potential to $r_c= 2^{\frac{1}{6}} \sigma$ to
cut the attractive tail. 

The spherical walls are concentric around the origin with radii of
$R_i = R - dR/2 - \sigma$ and $R_o = R + dR/2 + \sigma$, the shell
therefore has a thickness of $dR$, as shown in
fig.~(\ref{fig:simbox-sketch}).
The amount of curvature forced onto the system can then be tuned by
varying the radius of the spherical shell. The initial position of the
centers of the stars are chosen inside the simulation volume from a
uniform distribution. The arms are then placed at random positions
around the center. In order to achieve a well mixed configuration the
system runs $5.5 \times 10^{5}$ time steps where the interaction
parameter between any species of particles is set to
$a_{ii}=a_{ij}=25\,\frac{\mathcal{E}}{\mathcal{D}}$. After this warmup
phase the parameters are set as stated above according to their
species.  The temperature of the system was kept constant at $k_b T=1$
for the entire run. All simulations have been run with time steps of
$\Delta t=0.005$ and ran at least $3 \times 10^{8}$ time steps, larger
systems with $R>8\,\mathcal{D}$ ran $5 \times 10^{8}$ time
steps. After these long runs we assume that an equilibrium is reached,
which is confirmed visually in random samples.  The radius $R$ of the
spherical shell was varied with
$R=4,5,6,7,8,9,10\,\mathcal{D}$. Alternating the radius has two
effects: (1) due to constant a number density of
$\rho=3\,\frac{1}{\mathcal{D}^3}$ the number of molecules increase
with a larger shell volume; (2) the curvature of the shell decreases
with increasing radius. For each radius 20 configurations for each
$ABC$ and $ABB$ systems were simulated with different random
initialisations for statistical significance.

\subsection{Structure analysis of tiling patterns}

When the simulation is deemed to be equilibrated, the resulting
spherical tilings are recovered from the polymer configurations using
Set-Voronoi diagrams \citep{schaller2013} as implemented in
\textsc{pomelo} \citep{weis2017}. The aim is to substitute each domain
in the system with a polygon, representing a tile, where the number of
edges of the latter is equal to the number of neighboring domains.

In the $ABB$ systems we characterize the structure of the A beads,
considering the B particles as a matrix. We use a cluster algorithm
(implemented in the trajectory analysis package
\textsc{freud}\citep{freud}) to identify all A domains. This provides
a list of $N$ clusters, one for each $A$ type domain, each with a list
of which particles it is made from. Then the Voronoi diagram of all
$A$ type particles is computed, where all cells of particles belonging
to the same cluster are merged, leaving one cell per cluster. The
number of neighbors for each domain are determined from the number of
adjacent cells sharing a common edge. The spherical tiling is
recovered by representing each domain by a polygon which number of
edges equals the number of neighbors.

\begin{figure*}
  \centering
  \includegraphics[width=.9\textwidth]{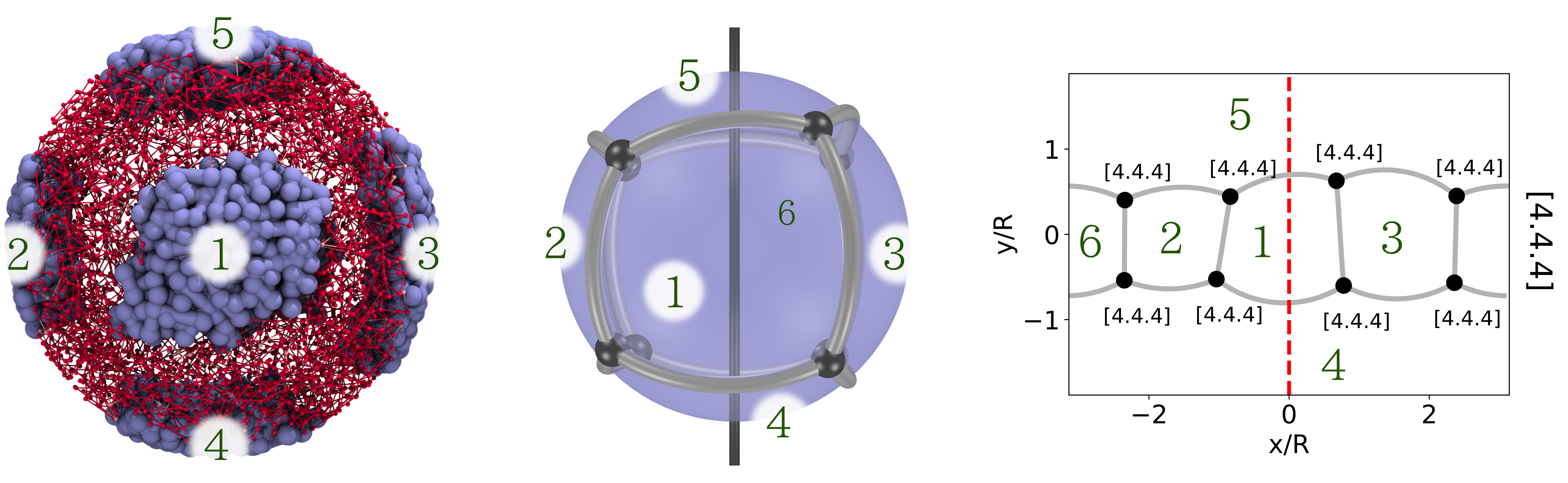}
  \caption{\textbf{Spherical tilings from polymeric self-assembly on a
      spherical substrate} \textit{Left panel:} A simulation snapshot
    of an equilibrated $ABB$ star copolymer melt confined to a spherical
    shell. The $A$-type arms have assembled into six domains, which
    arranged in a cuboidal symmetry in the $B$-type
    matrix. \textit{Middle panel:} The spherical tiling recovered from
    the polymer system. Each face in the tiling corresponds to a blue
    domain in the simulation snapshot as labeled. \textit{Right
      panel:} A modified Mercator projection of the spherical tiling,
    labeled with the corresponding tiles on the sphere and the domain
    in the simulation snapshot. Each vertex is labeled with its
    Schl\"afli symbol, the union of all types of distinctive vertex
    labels gives the Schl\"afli symbol of the entire tiling, as shown
    vertically on the far right. In order to be comparable, all
    tilings are rotated so all tilings of a type have the same
    orientation. The gray axis in the middle panel indicates the
    orientation of the red, dashed line in the projection in 3D.}
  \label{fig:analysis-abb}
\end{figure*}

For the visualizations of the tiling shown in
fig.~(\ref{fig:analysis-abb}) and (\ref{fig:tiling-abc}), the vertices
of a tiling are placed at the grafting points where three Voronoi
cells meet. The edges connecting vertices are great circle
segments. Figure~(\ref{fig:analysis-abb}) shows a simulation snapshot
on the left and a representation of the spherical tiling in the
middle. We obtain 2D topological representations of the tilings
through the Mercator projection as used in \citep{zhang2014}. Each
point on the sphere given in spherical coordinate angles $(\theta,
\phi)$ with $\theta \in [-\pi/2, \pi/2]$ and $\phi \in [0, 2\pi]$ is
mapped in the carthesian plane by $x=R \phi$ and $y = R \ln \left(
\tan \left( \frac{\pi}{4} + \frac{2}{5} \theta \right) \right)$.
An example of such a projection is shown on the right hand panel of
fig.~(\ref{fig:analysis-abb}). In these representations, the plot has
periodic boundary conditions in the x direction, however, not in y
direction. The top and bottom tiles therefore are not adjacent, but
represent the tiles at the poles of the sphere. The purpose of the
planar projections is to correctly capture the topology and neighbor
relations, not the geometry, which is deformed in the projection.

For the $ABC$ systems we analyse each of the three species
individually, treating the respective other two domains as the matrix,
and apply the same analysis as above. That is, to analyse $A$, we
consider $B$ and $C$ indistinguishable and to represent the matrix and
so on. From each $ABC$ system three tilings from the $A$, $B$ and $C$
type domains are obtained, as can be seen in
fig.~(\ref{fig:tiling-abc}).  Tilings are labeled using Schl\"afli
Symbols, see caption of table~\ref{tab:thomson-tilings} for
details. All simulation screenshots were made using the Tachyon render
\citep{stone1998} in \textsc{vmd} \citep{humphrey1996}, the renderings
of the tilings were created using a custom script in \textsc{blender}
\citep{blender}.

Every $1\times10^{6}$ simulation step, a snapshot of all particles was
made, which results in 300 frames over the simulation time. The
statistics shown in the result section include all tilings from the
latest 15\,frames of the simulation to account for invalid simulation
frames. Since 20 independent rubs were made for each radius, 300
frames were analysed for both the $ABC$ and $ABB$ systems. In the
$ABB$ system, this provides 300 spherical tilings, in the $ABC$ system
900 tilings, since there are three colors in each frame. Note,
however, that the ``real'' statistics are only based on 20 different
runs for each polymer type and radius, since the tiling in the last 15
frames of each run are assumed to be equilibrated and therefore is not
expected to change.\footnote[3]{In general the presented analysis
  method using Voronoi diagrams works well and is robust. In some rare
  cases, however, we find it to produce invalid results. These cases
  are, when very short edges appear in the Voronoi diagram, which
  means two vertices are very close together. In these cases the
  neighborhood relations are not clear for the algorithm and small
  displacements of a single particle can alter the resulting
  tiling. The other weak point is the cluster analysis. Since all
  systems are run at finite temperature, there might be particles
  moving outside their domain in the vicinity of another domains. The
  cluster algorithm then can mistake both clusters as a single
  one. Most of these invalid frames can be identified and then ignored
  by checking if $\chi \neq 2$. A neglectable number of
  $\approx1.5\,$\% frames were found to be invalid in the sense as
  discussed.}

\section{Results and discussion}

We find the following key features:

\begin{figure*}
  \begin{tabular}{cc}
    \begin{minipage}{0.43\textwidth}
      \includegraphics[width=\textwidth]{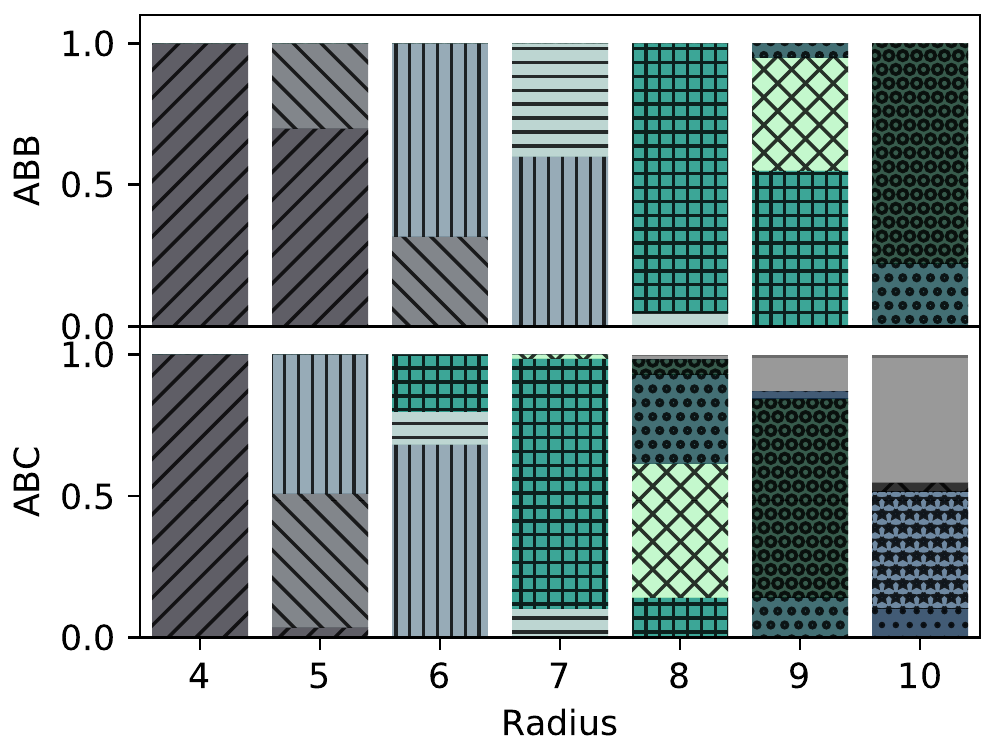}      
    \caption{ \textbf{Equilibrium configurations of self-assembly of
        star copolymers on a spherical substrate} Results of multiple
      runs of the self-assembly of $ABB$ (top) and $ABC$ (bottom) star
      copolymers on a spherical substrate of radius R. Each color
      denotes one type of tiling, as labeled in
      table~\ref{tab:thomson-tilings}. The plot shows the fraction of
      simulation snapshots found with the respective tiling for each
      radius. In general, multiple tilings are found as solutions for
      a single radius, where the results for $ABB$ and $ABC$ systems
      differ.}
    \label{fig:proportions-rad}
    \end{minipage}&  \raisebox{5pt}{
    \begin{minipage}{0.47\textwidth}
    \begin{tabular}{rrc||c||cccccc}
        &&&&\multicolumn{5}{c}{number of tiles with}\\
        &&N&Schl\"afli symbol&1&2&3&4&5&6\\
        &&&&\multicolumn{5}{c}{edges}\\
        \hline
        /&\crule[N2]&2&[1.1]&2&&&&&\\
        \textbackslash&\crule[N3]&3&[2.2.2]&&3&&&&\\
        |&\crule[N4]&4&[3.3.3]&&&4&&&\\
        -&\crule[N5]&5&[3.4.4]&&&2&3&&\\
        +&\crule[N6]&6&[4.4.4]&&&&6&&\\
        x&\crule[N7]&7&[4.4.5]&&&&5&2&\\
        o&\crule[N8]&8&[[4.4.5], [4.5.5]]&&&&4&4&\\
        O&\crule[N9]&9&[[4.5.5, 5.5.5]]&&&&3&6&\\
        .&\crule[N10]&10&[[4.5.5, 5.5.5]]&&&&2&8&\\
        *&\crule[N11]&11&[[4.5.5], [4.5.6],&&&&2&8&1\\
          &&& [5.5.5], [5.5.6]]&&&&&&\\    
        /\textbackslash&\crule[N12]&12&[5.5.5]&&&&&12&\\
        \hline
        &\crule[NT]&&Non-Thomson&&&&&&\\
        &\crule[IN]&&Invalid&&&&&&\\    
      \end{tabular}
      \captionof{table}{Solutions of the Thomson problem for systems
        with up to 12 electrons described as spherical tilings. The
        table shows the number of $n$-gons and the Schl\"afli symbol
        for the dual lattice of a solution of a $N$-electron Thomson
        problem. A Schl\"afli symbol \citep{kirkensgaard2014b,
          gemma2002, gruenbaum2013} is a set of $l$ numbers $[k_1,
          k_2, ... k_l]$ denoting that a vertex is adjacent to $l$
        tiles with $k_l$ edges respectively (see also right pane in
        fig.~(\ref{fig:analysis-abb})). The Schl\"afli symbol for an
        entire tiling just lists all different types of occurring
        vertices, see fig.~(\ref{fig:analysis-abb}). The symbols to the left relates to the textures in fig.~(\ref{fig:proportions-rad}).}
    \label{tab:thomson-tilings}
\end{minipage}}\\
\end{tabular}
\end{figure*}

\begin{figure*}
  \center
  \includegraphics[width=.9\textwidth]{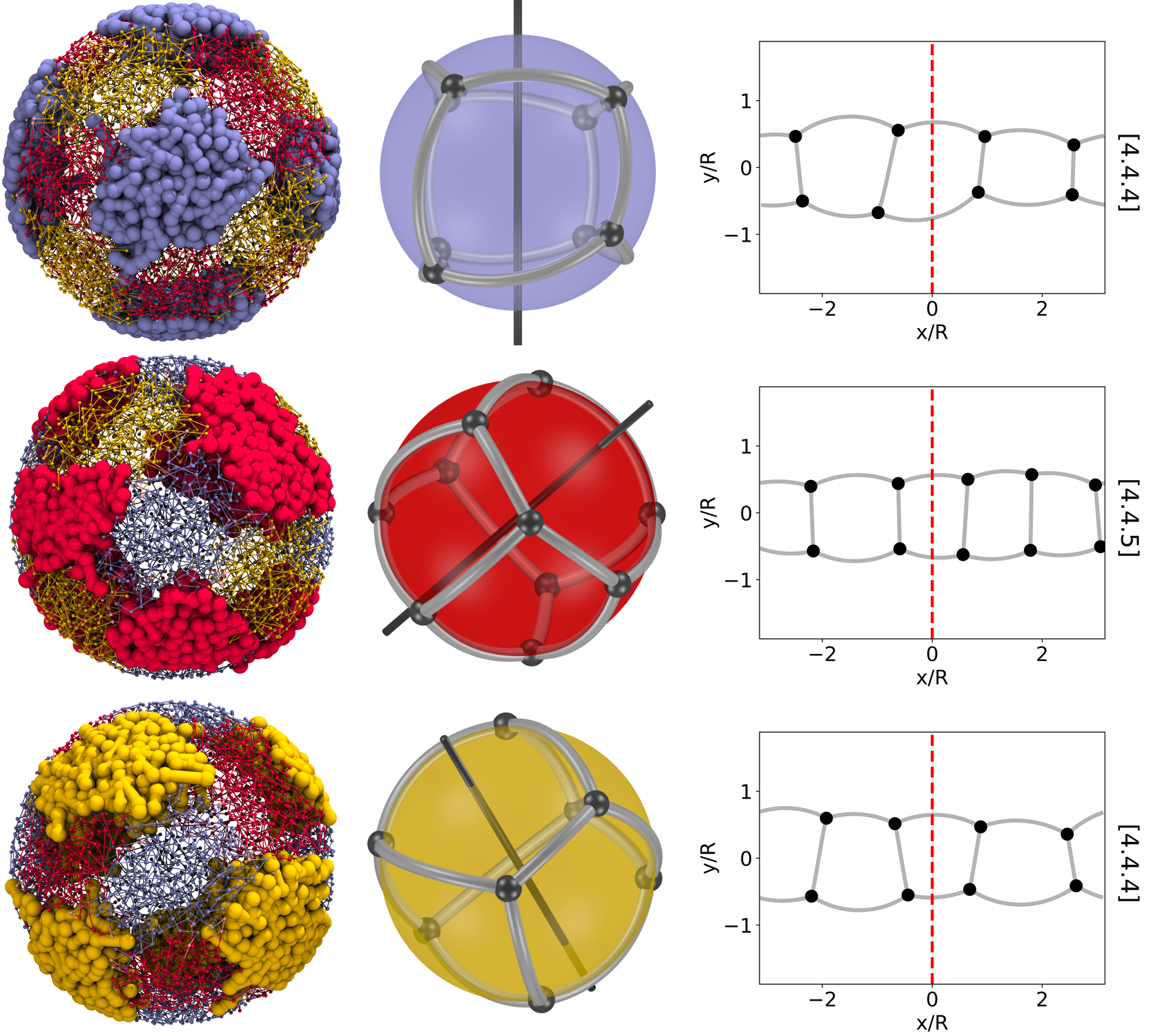}
\caption{\textbf{The three spherical tilings representing the
    structure of the A, B and C domains} The figure shows three
  tilings, generated by the self-assembly of $ABC$ star copolymers,
  in each of the colors on the same sphere. \textit{Left column:}
  Rendering of the simulation snapshot where the patches creating the
  tiling are emphasized while showing the others
  simultaneously. \textit{Middle column:} The tiling derived from the
  emphasized patch configuration on the left. The axis shows the
  orientation of the axis with the highest symmetry. \textit{Right
    column:} A 2D projection of the tiling. The black axis in the
  middle column indicates the position of the red line in the
  projection plots. The label of the tiling is given on the far right
  side.}
  \label{fig:tiling-abc}
\end{figure*}

\begin{itemize}
\item For spherical substrates with $R<8$ all tilings generated by
  both the $ABB$ and $ABC$ systems are identical to tilings generated
  from Thomson solutions, see fig.~(\ref{fig:proportions-rad}). Only
  for radii $R\geq 8$, we observed simulations of $ABC$ systems which
  were not Thomson-type solutions (see below). We only found Thomson
  solutions for the $ABB$ systems.
  
\item For $R>4$, instead of a single equilibrium solution, we find a spectrum of
  configurations, see fig.~(\ref{fig:proportions-rad}). Within our
  analysis, these appear as degenerate (or nearly degenerate)
  configurations that occur with statistical frequencies.
  
\item The analysis of the three colors of an $ABC$ tiling shows that
  they each individually form Thomson solutions, but not necessarily
  of the same tessellation type, see fig.~(\ref{fig:tiling-abc}).
  
\item The resulting tilings can be tuned by varying the radius of the
  sphere where the $ABC$ star copolymer system shows a different behaviour in
  the frequencies of the tilings than the $ABB$ star copolymer systems, see
  fig.~(\ref{fig:proportions-rad}).
\end{itemize}

\noindent To start our discussion we single out the $R=8$ systems to
illustrate the key results. Out of the 300 frames in the $ABB$ systems
with $R=8$, we find the majority of frames (about 95\,\%) to have 6
tiles in a $[4.4.4]$ configuration, only a very small proportions of
5\,\% has 5 tiles in a $[3.4.4]$ tiling. Both of the configurations
are identical to tilings generated by Thomson solutions. The $ABC$
case is slightly more complex: out of the 900 analysed tilings, we
find only about 14\,\% of the configuration with 6 tiles in a
$[4.4.4]$ tiling, 47\,\% with 7 tiles in a $[4.4.5]$ tiling, about
31\,\% with 8 tiles in a $[4.4.5, 4.5.5]$ tiling and 6\,\% with 9
tiles as a $[4.5.5, 5.5.5]$ configuration. As in the $ABB$ systems,
all of these are identical to tilings from Thomson solutions. Only
about 1\,\% of the tilings were found to differ from the Thomson
solution tilings.

While for $R=8$ the $ABB$ system overwhelmingly forms the same type of
tiling, the $ABC$ misses this feature. We find this behaviour across
most of the other systems on different radii: all of the $ABB$ systems
form at least two different tilings for each radius, the $R=9$ system
even three, however, all are Thomson solutions. All $ABC$ systems show
at least three different tilings for each radius, again almost all of
them are Thomson solutions. We find exceptions for the $R=4$ spheres,
where for both systems only a single type of tiling is found and the
$ABC$ system for $R=7$, where although three different tilings are
found, the majority (86\,\%) of analysed frames forms only one
type. Another exception are larger $ABC$ systems, where we find an
increasing number ($\approx$1\% for $R=8$, $\approx$12\,\% for $R=9$,
$\approx$44\,\% for $R=10$) of tilings not connected to the Thomson
problem. However, we do see these percentages go down as the
simulations are running for longer times so we conjecture that
eventually all non-Thomson solutions might anneal out.

Since the free energy levels of different tilings is a function of the
sphere radius, as will be discussed later, this may allow the
conclusion that the energy levels are almost degenerate for the
majority of the combinations of the chosen star copolymers and
simulation volume geometry. The finite temperature of our simulation
then allows the system to jump into local minima instead of the
energetic ground state, which results in the spectrum of tilings found
here. For some systems though ($R=4$, $ABB$ on $R=7$, $ABC$ on $R=8$) the
energy level of certain configurations seems to be deep enough to
prevent other structures to assemble.

For all of the $ABB$ system and $ABC$ systems with $R<8$ we find that
all tilings are of Thomson-type solutions, as seen in
fig.~(\ref{fig:proportions-rad}). This means that the $N$ $A$-type
(and respectively $B$- and $C$-type) patches will sit at the same
positions as the electrons would in a $N$-electron Thomson solution,
which we checked by visual observation. This is a remarkable result
since although the polymers only have short range interactions a
structure of long range order is formed. Such long range interactions
in a similar system of interacting micelles formed by diblock
copolymers has been predicted by \cite{semenov1985}.

For $ABC$ systems, with increasing radii beginning at $R \geq 8$ an
increasing number of configurations, up to $\approx 40$\,\% are not of
the Thomson-type. However, we argue that these systems are stuck in
local minima. The Thomson problem in general is known for its high
probability of being stuck in local
minima\cite{bondarenko2015}. Already for $N=32$ electrons the
probability of finding the global minimum drops significantly, which
roughly coincides with the threshold where our results show local
minima. Using polymer melts instead of point-like particles adds even
more complexity to the energy landscape. The fact that the $ABB$
systems - somewhat simpler than the $ABC$ system, due to the missing
color constraint for example - do not show any non-Thomson solutions
and that the proportion of non-Thomson solutions increase with
increasing radius, and thus number of patches, supports this.

Based on the observation that the simulations only show Thomson
solutions up to systems with $R=7$, all of those $ABC$ systems
must consist of three different tilings, each of which solves the
Thomson problem on its own while coexisting with two others on the
same sphere. We will elaborate on this again using an example system
from the $R=8$ runs using fig.~(\ref{fig:tiling-abc}), however, all
systems of various sizes share the same behaviour.

Figure (\ref{fig:tiling-abc}) shows how three different Thomson
solutions can coexist on a single sphere. When only considering the
blue patches, while red and yellow act as a matrix, the resulting
tiling is of type $[4.4.4]$. Analogue type $[4.4.5]$ and $[4.4.4]$
tilings are formed in the red and yellow patches respectively. The
tilings have a different orientation on the sphere as can be seen on
the orientation of the axis of highest symmetry.

In the case of the $R=8$ tilings, the overwhelming number of $ABC$
systems were found to be a combination of three Thomson solutions,
where the most prominent combinations are $(7/8/8)$ with 19\,\%,
$(7/7/8)$ with 15\,\%, $(7/7/7)$ with 15\,\% and $(6/7/7)$ with
14\,\%. The number denotes the number of patches in the tiling, the
tiling label can be found using table~\ref{tab:thomson-tilings}. Only
a small number of 1\,\% of the systems contained tilings which are not
Thomson solutions. The same holds for systems with smaller radius:
since no tilings were found not to solve the Thomson problem, all
combinations there consist of Thomson solutions. For larger systems
the number of non-Thomson configuration increases.  Therefore, the
amount of configurations containing non-Thomson solutions increases as
well, however, we cannot say if that is caused by increasingly
difficult equilibration or due to other reasons, for example the color
constraint preventing some combinations to be assembled.

In the $R=8$ case we found the $(7/7/7)$ combination formed in two
different ways: once with only Thomson solutions, and once with one
Thomson problem configuration and two non-Thomson tilings. In this
case we clearly see that it is possible to build this $(7/7/7)$
combination using only Thomson solutions. This hints towards our guess
that combinations with non-Thomson solutions are not caused by the
color constraint but equilibration issues. Based on the data and this
assumption we suspect that the color constraint on its own does not
influence the resulting tiling. Apart from this observation we could
not find any regularities in the frequencies of the different
combinations of the tilings for any radius.

Having said that the color constraint seems to not have any effect on
the resulting tilings, it is important to note that although the
simulation data of $ABB$ systems for $R=8$ clearly shows the type
$[4.4.4]$ tiling as the minimal energy configuration, we could only
find one out of 300 combinations being a $3 \times [4.4.4]$ tiling
on a single sphere in the $ABC$ systems of the same radius. Therefore
this finding cannot be a result from the imposed color constraint but
must rather be caused by a change in the energy functional.

This observation is found as general behaviour across all runs: there
are statistically significant differences in the frequency with which
tilings occur in $ABB$ and $ABC$ systems. Therefore, the self-assembly
process of ABB and ABC does show differences that are manifested in
the topology of the adopted structures, or at least in the statistical
properties of these properties. This observation enables us to answer
the following question: if one is only able to see one kind of color,
for example by looking through colored filters, is it possible to
determine if the observed structure is assembled by an $ABB$ or an
$ABC$ system? The answer depends on the circumstances: if only one
configuration is available, the answer is no, since each configuration
found in this article is assembled by both the $ABB$ and the $ABC$
systems. However, if multiple samples are available, the answer is
yes. The different statistics of frequencies of the tilings in the
different systems as shown in fig.~(\ref{fig:tiling-abc}) enables us
to determine the type of systems of the given samples.

When assuming that all equilibrium solutions are Thomson solutions,
the resulting tiling type is determined by the number of patches
assembled in the melt. As can be seen in the data, this number is a
function of the radius: with increasing sphere size the number of
patches increases. This follows from the energy functional of polymer
melts in the strong segregation limit:
\begin{equation*}
  F = F_{\text{conf}} + F_{\text{int}}
\end{equation*}
where $F_{\text{conf}}$ is the entropic contribution determined by the
domain shape and $F_{\text{int}}$ is the enthalpic contribution and
measures the interface area \cite{semenov1985}. The entropic
contribution favours most spherical patch shapes and penalises domains,
where the polymer chains have to be stretched. Consider an exemplary
$ABB$ system with two $A$-type patches located at the north and south
pole of the sphere. The interfaces, on which the grafting points of
the star copolymers must sit, are then disk segments centred around the
poles. From there the $B$ type arms stretch to cover the entire
sphere. Since the arm length is kept constant, the arms must stretch
increasingly with increasing sphere radius to cover the surface of the
sphere. This comes with an entropic penalty. At some point it becomes
energetically more favorable to change to a three patch configuration
which, despite the increasing interface energy, relaxes the polymer
arms and reduces the entropic energy contribution. This interplay
between minimising the interface area and entropic energy contribution
determines the number of domains and, thus, the resulting structure in
these systems.

The differences in the frequencies of tilings in $ABB$ and $ABC$
systems are caused by a modification of the energy functional. While
in the $ABB$ systems only $A-B$ interfaces exist, the $ABC$ systems
also develop $A-C$ and $B-C$ interfaces, thus, increasing the
enthalpic energy contribution. The entropic contribution changes since
the grafting points of the star copolymers are not allowed to move
freely along the $A-B$ interface but are constrained to $ABC$ triple
lines, resulting in more constrained polymer paths leading,
presumably, to an entropic penalty.

\section{Conclusion and Outlook}

In this article the self-assembly of $ABC$ and $ABB$ star copolymers
confined to a spherical shell was simulated using DPD molecular
dynamics simulations in order to investigate the combined influence of
geometric frustration and the color constraint inherent in the $ABC$
system. In bulk simulations, these polymers form columnar phases whose
cross sections are 3-colored, planar, hexagonal tiling patterns. The
architecture of $ABC$ stars imposes the color constraint onto the
resulting structures: only tiles with an even number of edges are
allowed where the color of all adjacent tiles must alternate. To
differentiate between the influence of the color constraint and the
curvature $ABB$ systems were simulated as reference, where the color
constraint does not apply.

We can summarise our findings into four core results: (1) apart from
kinetically stuck configurations in large $ABC$ systems $(R\geq 8)$,
we find all tilings in both $ABB$ and $ABC$ systems to be Thomson
solutions. (2) In $ABC$ systems we find three possibly different
tilings on each sphere, one in each color while neglecting the other
two, all of which solve the Thomson problem for small radii
individually. We find some non-Thomson solutions for larger radii ($R
\geq 8$) but believe this is due to equilibration issues as discussed
above.

(3) A spectrum of configurations dominates the ensemble, rather than a
single structure. This leads to the occurrence of a small number of
different tilings in both $ABB$ and $ABC$ systems, with varying
probability. (4) The latter can be tuned by varying the sphere radius,
which means we can switch between Thomson solutions of different
numbers of particles. While we could not find any combination of three
tilings on the same sphere which the color constraint does not allow,
the frequencies of the tilings in the $ABC$ compared to the $ABB$
system show statistically significant differences.
This provides the possibility to differentiate between the two
systems: statistically speaking we can determine if a tiling was
formed by an $ABB$ or and $ABC$ system by only being able to see a
single color.

A direct comparison of our work to the presented results of frozen
particles on a sphere \citep{bowick2000} or the diblock copolymers on
a spherical substrate \cite{zhang2014} is somewhat difficult: while in
these systems the number of particles is in the order of 100 (which is
equivalent to one patch in our systems), our largest system consists
of a maximum of 12 domains in a single tiling. At these system sizes
the tilings are missing the regularity to define ``defects'' in their
structure.
Taking the entire $ABC$ system with all of its colors into account,
however, we have a system consisting of up to 37 patches. Here we can
see the influence of the color constraint: instead of finding isolated
pentagonal defects or scars of pentagons and heptagons, we find either
six squares or a combination of squares and octagons to cope with the
geometrical constraint.

A promising and interesting application of this work can be found in
the field of patchy particles: discrete particles with patches on
their surface which can couple and form bonds to other patches,
e.g. Janus-particles \citep{zhang2004_patchy, casagrande1989}. A
mechanism to assemble such particles and tune their coordination
number include self-assembly of monolayers of surfactants on spherical
substrates \citep{ponssiepermann2012a, ponssiepermann2012b}. Our work
presents an example how such a self-assembly could be
realised. Instead of using two repulsive walls to model a spherical
substrate, the architecture and composition of the polymer can be
modified, so that spherical droplets will form in solution, as seen in
previous work\citep{higuchi2008a,higuchi2008b, jeon2007,
  ku2019}. Preliminary simulations showed that adding a fourth arm,
immiscible with the already existing arms, would form the core of such
a droplet, on which surface the $ABC$ arms assemble as presented in
this work. By tuning the length of this fourth arm, the radius of the
droplets can be changed. As shown in this article, the number of
patches would change and therefore the coordination number of the
droplet as a patchy particle. Further, instead of using symmetric star
copolymers, the length of one arm can be varied to generate different,
'asymmetric' tilings, analogue to \citep{kirkensgaard2014a,
  gemma2002}.

\section*{Conflict of Interest}
The Authors declare no conflict of interest.

\section*{Acknowledgment}
This work was supported by resources provided by the Pawsey
Supercomputing Centre with funding from the Australian Government and
the Government of Western Australia, as well as LUNRAC at Lund,
Sweden.  We like to thank Nigel Marks from Curtin University, Perth,
for fruitful discussions about the Thomson problem.

\balance

\renewcommand\refname{References}

\bibliography{phd_refs.bib} 
\bibliographystyle{rsc} 

\end{document}